\documentclass[usenatbib,usedcolumn]{mn2e}
\usepackage{psfig,epsf,amsmath}

\newbox\grsign \setbox\grsign=\hbox{$>$}
\newdimen\grdimen \grdimen=\ht\grsign
\newbox\laxbox \newbox\gaxbox
\setbox\gaxbox=\hbox{\raise.5ex\hbox{$>$}\llap
        {\lower.5ex\hbox{$\sim$}}}\ht1=\grdimen\dp1=0pt
\setbox\laxbox=\hbox{\raise.5ex\hbox{$<$}\llap
        {\lower.5ex\hbox{$\sim$}}}\ht2=\grdimen\dp2=0pt
\newcommand{\ion}[2]{\mbox{\textrm{#1}$\;$\textsc{#2}}}
\newcommand{\hei}{\mbox{He$\,${\sc i}}}
\newcommand{\kms}{\mbox{${\;{\rm km\,s^{-1}}}$}}
\newcommand{\Msolar}{\mbox{${\; {\rm M_{\sun}}}$}}

\title{The nature of the companion star in Circinus X-1}
\author[H. M. Johnston, R. Soria and J. Gibson]{Helen
M. Johnston$^1$\thanks{E-mail: H.Johnston@physics.usyd.edu.au},
Roberto Soria$^2$ and Joel Gibson$^1$ \\ 
$^{1}$SIfA, School of Physics, University of Sydney, NSW 2006, Australia\\
$^{2}$ICRAR, Curtin University, Bentley, WA 6102, Australia}

\date{Received: }

\begin{document}

\maketitle

\label{firstpage}

\begin{abstract}

We present optical spectra and images of the X-ray binary Circinus
X-1. The optical light curve of Cir X-1 is strongly variable, changing
in brightness by 1.2 magnitudes in the space of four days. The shape
of the light curve is consistent with that seen in the 1980s, when the
X-ray and radio counterparts of the source were at least ten times as
bright as they are currently. We detect strong, variable H$\alpha$\
emission lines, consisting of multiple components which vary with
orbital phase. 
We estimate the extinction to the source from the
strength of the diffuse interstellar bands and the Balmer decrement;
the two methods give $A_V = 7.6 \pm 0.6\;\mathrm{mag}$ and $A_V >
9.1\;\mathrm{mag}$ respectively. The optical light curve can be
modelled as arising from irradiation of the companion star by the
central X-ray source, where a low-temperature star fills its Roche
lobe in an orbit of moderate eccentricity ($e \sim 0.4$). We suggest
that the companion star is over-luminous and under-dense, due to the
impact of the supernova which occurred less than 5000~yr ago.

\end{abstract}



\section{Introduction}
\label{sec:intro}

Cir X-1 is a highly unusual X-ray binary, whose nature has been a
puzzle for many years. A 16.6~d period was discovered in X-ray
modulations \citep{khbs76}, and \citet{wmw+77} discovered radio flares
with the same period. The X-ray intensity of the source has changed by
two orders of magnitude over the more than 40 years since its
discovery, ranging from a peak of twice the intensity of the Crab
nebula in the late 1990s \citep{stb+03} to 10 mCrab in 2008--2009
\citep{nmk+10}. Since 2010, the source has begun to brighten again at
X-ray wavelengths, suggesting it may once more be moving into an
active phase.

The accreting star was shown to be a neutron star by the discovery of
Type~I X-ray bursts observed in a brief episode in 1984--1985
(Tennant, Fabian \& Shafer 1986\nocite{tfs86b}).  During this period
of very low X-ray activity, the X-ray timing behaviour put Cir X-1 in
the class of \textit{atoll} sources \citep{okk+95}, believed to be
neutron stars with low magnetic fields accreting at a low
rate. However, \textit{RXTE} observations when the source was brighter
showed clear Z-source behaviour \citep{sbl99}.  Z-source behaviour is
characteristic of neutron stars accreting near the Eddington limit
from a low-mass companion; Cir X-1 is unusual in that it only
sometimes shows this behaviour. X-ray bursts were detected in 2010 as
the source began to brighten again \citep{lwa+10}; these bursts were
the first seen for 25 years since the initial bursting episode in
1984--1985 \citep{tfs86b}.

With all these conflicting characteristics, the nature of the
companion star remains extremely unclear.  The optical counterpart was
identified as a faint red star with strong H$\alpha$\ and weak \hei\
lines \citep{wmw+77}. From the strength of the $\lambda$6284 diffuse
interstellar absorption line, \citet{wmw+77} found a rather uncertain
estimate for the reddening of the optical counterpart of $E(B-V) \sim
3.5$~mag.  The large reddening, combined with the eccentric orbit
evidenced by the X-ray modulation, suggested that the companion star
was an early-type supergiant transferring mass via a stellar wind.

\citet{mjh+80} suggested a model for the system where the X-rays are
periodically absorbed by a stellar wind from a massive O or B
supergiant ($M \simeq 20 \Msolar$) in a highly eccentric orbit, $e
\sim 0.7$ to 0.8. This model, and any other model involving a massive
star as the mass donor, require large values for the reddening, $A_V
\sim 11$~mag.

Doubts were cast on the massive star model by subsequent observations.
\citet{nfg80} found correlated changes in the optical and infrared
fluxes from Cir~X-1, which suggests that they are physically related.
Further, the large drop in both fluxes between 1976 and 1979 suggests
that the donor star cannot be dominating the light from the system,
and the large changes in brightness must be related to the infalling
material. \citet{as82} showed that the optical counterpart had been
wrongly identified in the very crowded field, and was in fact several
magnitudes fainter than that measured by \citet{wmw+77}. This was
further clarified by \citet{mon92}, who found that the counterpart is
in fact one of three (unrelated) stars within 1\farcs5.  \citet{hay87}
suggested the donor star was more likely to be a main-sequence B-star,
with $M \sim 5\Msolar$. \citet{snp+91} concluded, from the revised
magnitude and the interstellar absorption derived from X-ray
observations, that $A_V \ga 3.0$ and the donor star is a main
sequence star with spectral type no earlier than type G.
\citet{jnb07} detected near-IR Paschen absorption, which they
attributed to stellar lines from a supergiant B5--A0 star.


Recently, \citet{hsf+13} detected a young supernova remnant around
Circinus X-1 in X-ray and radio observations. Assuming this is the
remnant from the supernova that gave rise to Cir X-1's neutron star,
this places an upper limit of $t < 4600$~yr on the age of the system,
which would make Cir X-1 the youngest known X-ray binary. This very
young age places further constraints on the companion star; if it is
low-mass, it must still be on the main sequence, as there has not been
enough time for significant evolution since the supernova event.

Unfortunately, efforts to unambiguously detect the companion star
itself at optical wavelengths have been unsuccessful. The visible
spectrum is dominated by strong, asymmetric H$\alpha$\ emission
\citep{mcb97}.  \citet{jfw99} and \citet{jwfc01} found this H$\alpha$
line to be highly variable, both over a single orbit, and over many
years. It is not clear where the H$\alpha$\ emission is produced, and
since it consists of several components, it may arise in more than one
region in the binary.  No stellar features were identified in the very
red spectrum.

The extinction towards the source is likewise uncertain; the unknown
nature of the companion affects any estimate of the extinction.  If
the companion star is massive, then the extinction to the source must
be large, in order to produce the observed red spectrum. A late
spectral type companion, on the other hand, would require less
reddening to match the observed colours.  

Different authors have used X-ray observations to determine the
interstellar absorption towards the source, with varying
results. \citet{snp+91} suggest the observed low-energy cut-off gives
$N_H \sim 6 \times 10^{21}\;\mathrm{cm}^{-2}$, giving $A_V \sim 3$,
using the relation of \citet{ps95}: $N_H = 1.79 \pm 0.03 A_V
\;[\mathrm{mag}] \times 10^{21} \;[\mathrm{cm}^{-2}]$. Fits to data
from ROSAT and ASCA converge on a value of $N_H \sim 2 \times
10^{22}\;\mathrm{cm}^{-2}$ \citep{mcb97}, which implies $A_V \sim
11$. RXTE and Swift spectra \citep{abp+12} give a minimum value for
the interstellar absorption of $N_H \sim 1 \times
10^{22}\;\mathrm{cm}^{-2}$, which corresponds to $A_V >
5.6\;\mathrm{mag}$. 
These X-ray derived columns are effectively upper
limits on the interstellar contribution to the absorption. 

The distance to the source was recently determined from an X-ray light
echo to be $9.4^{+0.8}_{-1.0}\;\mathrm{kpc}$\ \citep{hbb+15}, a
distance consistent with previous best estimates of around
$8\;\mathrm{kpc}$ (see \citealt{hsf+13} for a discussion of the various
suggested distances to the source).

In this paper, we report on spectroscopic observations obtained in
order to determine the spectral type of the companion. In
Section~\ref{sec:obs-data-reduct} we describe our Gemini observations;
in Section~\ref{sec:photom} we discuss the light curve of the source,
and in Section~\ref{sec:spectroscopy} we put limits on the extinction
towards the source. 
In Section~\ref{sec:discussion} we discuss the
constraints that our observation put on the nature of the companion,
and a possible model for the system that satisfies these constraints.

\section{Observations and data reduction}
\label{sec:obs-data-reduct}

We used the Gemini Multi-Object Spectrograph \citep[GMOS; ][]{hja+04} on
the 8~m Gemini-S telescope to observe Cir X-1 on six separate
occasions in February--April 2013; a complete log of observations is
shown in Table~\ref{tab:obs-log}. Observations were carried out in
queue-scheduled mode.  The R400 grating was used in conjunction with
the G5325 blocking filter, giving a dispersion of 1.36~\AA/pixel and a
resolution of 8 \AA\ FWHM (measured from arc lines). The CCD was
binned by 2 in the spatial direction for the spectral exposures,
giving a pixel scale of 0.146~arcsec/pixel. Severe fringing affects
the red end of each spectrum, making the region beyond $7200\;$\AA\
unusable. 

Acquisition images were taken with each observation using an $r$-band
filter (r\_G0326); the exposure time for the acquisition images was
10~s.  The seeing was measured from the FWHM of the stars on the
acquisition images, and is listed in Table~\ref{tab:obs-log}.

The field is crowded, with two nearby unrelated stars within 1\farcs5
\citep{mon92}, so a slit-width of 1.0 arcsec was used, and the slit
was oriented 8\fdg5 east of north. This put star 2 of \citet{mon92} on
the slit as well as Cir X-1. A seeing condition (seeing $< 0\farcs8$)
was set on the scheduling of the observations to ensure that the two
objects were well-separated on the slit. No constraint was put on sky
transparency, so the data are not photometric.

\begin{table}
\caption{Journal of observations of Cir~X-1. Columns show the ID of
  the spectrum, the date of the observation, the modified Julian date
  of the centre of each observation, the exposure time, and FWHM
  seeing. The last two columns show the phase of the mid-point of the
  observation, according the quadratic ephemeris of Nicolson (2007),
  and the cycle number $N$ from this ephemeris, being the number of
  16.6~d cycles since MJD=43076.32. }
\label{tab:obs-log}\nocite{nic07}
\addtolength{\tabcolsep}{-1.5pt}
\begin{tabular}{llccccc}
\hline
& UT Date & MJD & $t_{\mathrm{exp}}$ & \multicolumn{1}{c}{Seeing} & Phase & Cycle \\
&         &     & (s)                & \multicolumn{1}{c}{($''$)} & $\phi$ \\ 
\hline
A & 2013 Feb 22 & 56346.338 & 6300 & 0.6 & 0.018 & 802 \\
B & 2013 Feb 23 & 56347.381 & 2100 & 0.8 & 0.081 & 802 \\
C & 2013 Feb 26 & 56350.304 & 3150 & 0.7 & 0.257 & 802 \\
D & 2013 Mar 02 & 56354.259 & 1050 & 0.9 & 0.496 & 802 \\
E & 2013 Mar 18 & 56370.237 & 4200 & 0.9 & 0.464 & 803 \\
F & 2013 Apr 21 & 56404.213 & 2100 & 0.8 & 0.521 & 805 \\
\hline
\end{tabular}
\end{table}

The data were reduced using standard GMOS tools in {\sc iraf}. Each
observation block consisted of series of 1050~s exposures of our
source, interspersed with flat-fields, with the central wavelength
alternated between 6500~\AA\ and 6550~\AA\ in order to eliminate the
small wavelength gaps between the detectors. During the reduction, all
the spectra from the same night were added together to create a single
spectrum with total exposure time as indicated in
Table~\ref{tab:obs-log}. We named these six spectra A--F, as shown in
the first column.  Standard reduction procedures were followed to
remove bias and pixel-to-pixel gain variations, extract the spectra
and determine the wavelength solution.  Flux calibration was performed
by comparing with the spectrum of the observed flux standard LTT~6248.
However, since the data were taken in non-photometric conditions, the
flux calibration should be considered only approximate.

The second last column of Table~\ref{tab:obs-log} shows the phase of
each observation, calculated from the ephemeris of \citet{nic07},
based on the timing of the radio flares. The phases of the spectra
range between $0.02$ (just after phase 0, which we take to represent
periastron) to $0.52$ (just after apastron).  The good period coverage
of the observations (at least of the first half of the orbit) was
serendipitous, an accidental outcome of queue scheduling with a good
seeing constraint.

\section{Photometry} \label{sec:photom}

\begin{figure}
     \centerline{\psfig{figure=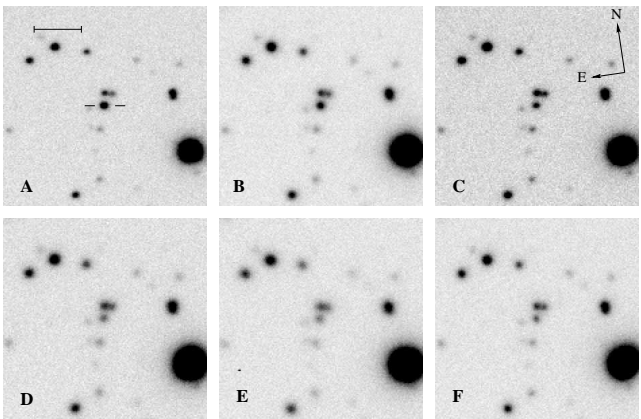,width=\columnwidth,clip=t}}

     \caption{Montage of Cir X-1 in each of the six acquisition images
     (Table~\ref{tab:obs-log}). Cir X-1 is indicated by the horizontal
     bars in the top left panel; the scale bar in the top left corner
     is $5''$ in length. North is up and east is to the left, skewed
     by 8\fdg5 E of N, as shown in the top right
     panel.}\label{fig:montage}

\end{figure}

Inspection of the acquisition images (Figure~\ref{fig:montage}) shows
that the brightness of Cir X-1 changed dramatically over the six
epochs that we have data for.  We used these $r$-band acquisition
images to investigate the changing brightness of the source.  Although
the data were not taken under photometric conditions, the field is
crowded, so we were able to do relative photometry with
ease. Instrumental magnitudes were determined using the aperture
photometry function of the \texttt{gaia} (Graphical Astronomy and
Image Analysis) tool. Seven reasonably isolated stars in the vicinity
of Cir X-1 were chosen, including the stars used by \citet[][see his
Fig. 1]{mon92}. The zero-point was estimated using a nearby star from
the USNO-B1.0 catalogue.

\begin{figure}
     \centerline{\psfig{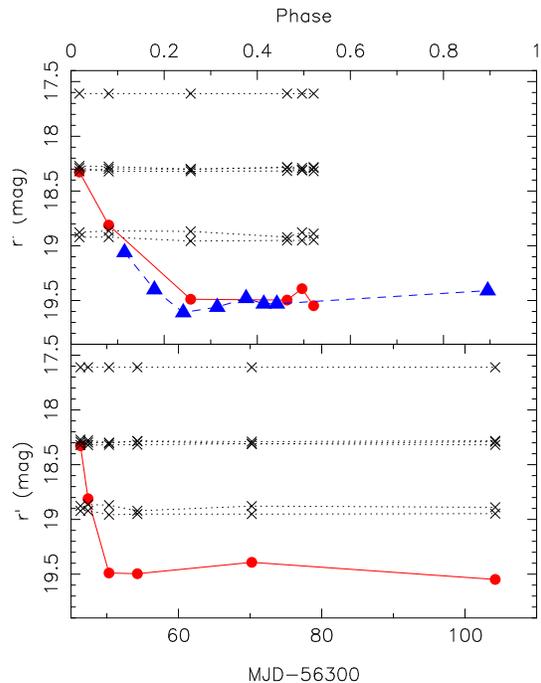}}

     \caption{Instrumental $r$-band light curve for Cir X-1 (filled
     circles, red solid line), relative to several field stars
     (crosses, black dotted line). The brightest object is USNO-B1.0
     0328$-$0672991, which was used to determine the (approximate)
     zero-point. The light curves are shown as a function of phase
     (top panel) and of MJD (bottom panel). The blue open triangles
     (dashed line) show the $R$-band light curve of Moneti (1992),
     with re-calculated phases (see text). }\label{fig:phot}

\end{figure}

We measured the brightness of Cir X-1 on the six acquisition images,
taken on the dates listed in Table~\ref{tab:obs-log}. The results are
shown in Figure~\ref{fig:phot}, as a function of both phase and MJD.
The light curve shows a drop of 1.2 magnitudes over a space of 4 days
($\sim 0.25$ in phase) during February 2013.

We cannot tell \textit{a priori} if this is an orbital change or a
secular change in brightness, since we have very sparse time coverage.
Comparison with the only previous study of the optical light curve of
Cir X-1 is instructive, however. The shape of the change in brightness
that we see is similar to, though larger than, that observed by
\citet{mon92}, in data from 1989. He reported a drop of 0.6 mag in V
and R between phase 0.07 and 0.2, followed by a slow rise at R up to
phase 0.8. We note that if we re-calculate the phases of Moneti's data
points using the ephemeris of \citet{nic07}, the two light curves line
up almost exactly (see Figure~\ref{fig:phot}); the larger change in
brightness seen in our photometry is then due to the fact that we have
a data point closer to the time of the radio flare.

It seems reasonable to conclude, then, that we are seeing variations
directly linked to the orbit in the brightness of Cir X-1. This
implies that the optical light curve has changed suprisingly little
over the past 25 years, despite the X-ray intensity, radio flux
density, and H$\alpha$ equivalent width changing by factors of 10 or
more over that time \citep[see e.g.][]{afn+13,jfw99}.

\subsection{Comparison with X-ray light curve}
\label{sec:X-ray-lc}

\begin{figure}
     \centerline{\psfig{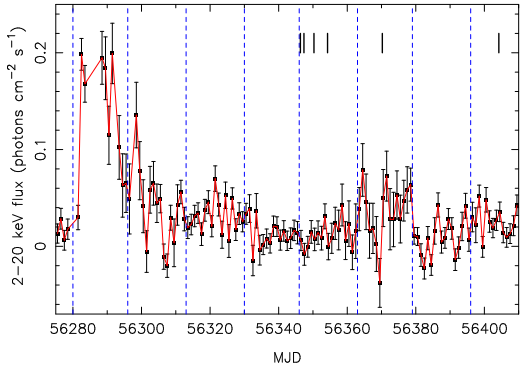}}
     \caption{X-ray light curve from MAXI, over the whole 2--20~keV
     band. The dashed lines show the times of periastron; the thick
     black vertical bars show the times of our optical spectra
     (Table~\ref{tab:obs-log}). The optical spectra were taken during
     a period of low X-ray activity.}\label{fig:Xray-lc}
\end{figure}

\citet{abp+12} found that since 2007, Cir X-1 has been in a state
characterised by persistent mCrab X-ray flux, punctuated by sporadic
several-week-long outbursts.  In Figure~\ref{fig:Xray-lc} we show the
X-ray light curve from the Monitor of All-sky X-ray Image
\citep[MAXI;][]{mku+09}. During the times of our optical spectra
(indicated by black vertical bars), the X-ray activity of the source
was low; there had been an outburst several orbits earlier, but the
average X-ray flux during our observations was $0.02~\mathrm{photons}
\,\mathrm{cm}^{-2}\,\mathrm{s}^{-1}$. 
Assuming a power-law spectrum
with photon index $\Gamma = 1.7$, this corresponds to a an X-ray flux
(0.1--100~keV) of $9.6 \times
10^{-10}\;\mathrm{erg}\,\mathrm{cm}^{-2}\,\mathrm{s}^{-1}$, or a
luminosity $L_X = 1.0 \times 10^{37}\;\mathrm{erg}\,\mathrm{s}^{-1}$,
assuming $N_H = 2 \times 10^{22}\;\mathrm{cm}^{-2}$ and $d =
9.4\;\mathrm{kpc}$.

Because of the faintness of the source during the period when our
spectra were taken, there is no X-ray spectrum available.

\section{Spectroscopy}
\label{sec:spectroscopy}

The optical spectrum of Cir X-1 consists of an extremely red
continuum, with several emission lines: H$\alpha$ $\lambda$6563, \hei\
$\lambda$6678, and \hei\ $\lambda$7065. Several clear absorption
features are also visible, but these are all due to diffuse
interstellar bands; see Section~\ref{sec:DIBs}. 
No stellar features
are visible, with the possible exception of a single unidentified line
near 6980 \AA. This line does not appear to be associated with a DIB,
and is visible in all but one of the six epochs, as can be seen in
Figure~\ref{fig:allspec}. 

\begin{figure}
     \centerline{\psfig{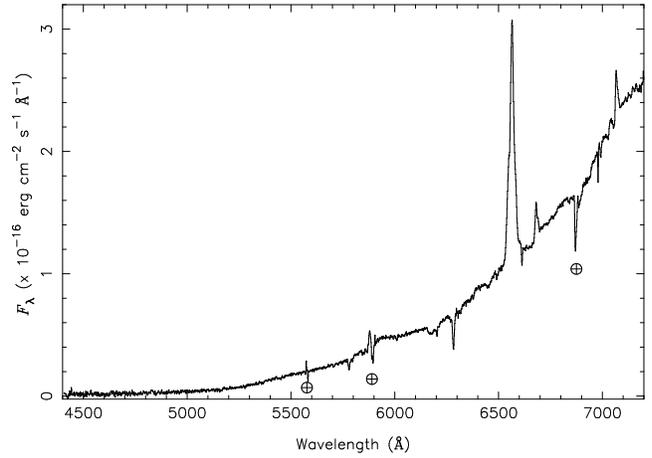}}
     \caption{The highest signal/noise spectrum of Cir X-1, spectrum
     A, taken on 2013 Feb 22. The spectrum is the sum of six separate
     exposures. The continuum is extremely red; emission lines of
     H$\alpha$ $\lambda$6563, \hei\ $\lambda$6678, and \hei\
     $\lambda$7065 can be seen, as well as absorption features, which
     are nearly all due to interstellar
     absorption.}\label{fig:spectrum}
\end{figure}

Figure~\ref{fig:spectrum} shows the spectrum with the highest S/N,
spectrum A.

\subsection{Emission lines}
\label{sec:emission}

\begin{figure}
     \centerline{\psfig{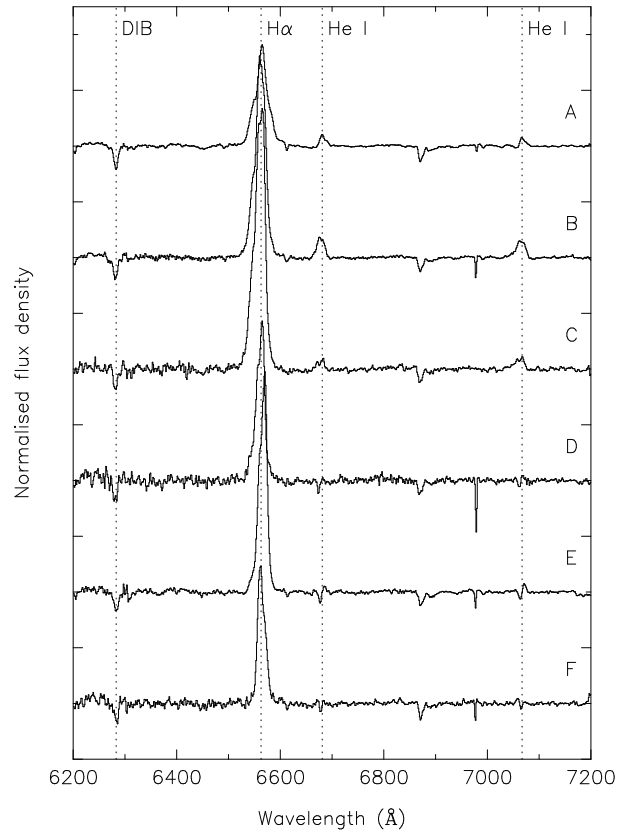}}
     \caption{The region around H$\alpha$ in all six epochs; the name
     of each observation is shown at the right of each spectrum, which
     are offset vertically for clarity. Each spectrum has been
     normalised by a polynomial fit to the continuum.  The position of
     the H$\alpha$ and \hei\ emission lines, as well as the strongest
     of the diffuse interstellar bands, are shown with dotted lines;
     see Fig.~\ref{fig:DIBs} for the identification of other
     features.  The flux of the H$\alpha$ line is greatest in spectrum
     A, but the equivalent width is largest in spectrum C because of
     the fading of the continuum between the two epochs
     (Table~\ref{tab:Ha}). }\label{fig:allspec}
\end{figure}

All six spectra show a strong H$\alpha$ emission line, but both the
shape and intensity of the line vary significantly between the
observations.  Much weaker \hei\ lines are also
visible. Figure~\ref{fig:allspec} shows the region around H$\alpha$
containing the emission lines in all six spectra.

\subsubsection{H$\alpha$ brightness}

The equivalent width (EW) of the H$\alpha$\ line ranges
from 40--120$\;$\AA\ in the different spectra. Because the continuum level
is changing dramatically over the six epochs (\S~\ref{sec:photom}),
some of this variation in EW is due to the changing brightness of the
continuum. Table~\ref{tab:Ha} shows the line flux and equivalent width
of the H$\alpha$\ line in each spectrum, measured by direct summation
of pixels in the line. 

\begin{table}
\caption{Variability in the H$\alpha$\ emission lines. The columns
show the name and phase $\phi$ of the spectrum, and the equivalent
width $W$\ and the flux of the H$\alpha$\ line.}
\label{tab:Ha}
\begin{tabular}{cccc}
\hline
         & & $W$   &  $f_\mathrm{H\alpha}$          \\ 
Spectrum & $\phi$ & (\AA) & ($\times 10^{-15} \; \mathrm{erg}\,\mathrm{cm}^{-2}\,\mathrm{s}^{-1}$) \\
\hline
A & 0.018 &  $46 \pm 2$ & $5.3 \pm 0.2$ \\
B & 0.081 &  $89 \pm 6$ & $5.0 \pm 0.3$ \\
C & 0.257 & $117 \pm 7$ & $3.5 \pm 0.2$ \\
D & 0.496 &  $54 \pm 6$ & $1.2 \pm 0.1$ \\
E & 0.464 &  $70 \pm 4$ & $1.9 \pm 0.1$ \\
F & 0.521 &  $40 \pm 9$ & $0.9 \pm 0.2$ \\
\hline
\end{tabular}
\end{table}

In \citet{jwfc01}, we reported that the EW of the H$\alpha$\ line
declined from 580$\;$\AA\ in 1976 to 10--20$\;$\AA\ in 2000. Subsequent
observations found that EW of the line continued to decline; by 2006
it had dropped to $\la 5\;$\AA, and was undetected on some occasions
\citep[][in prep.]{jws15}. The brightness of the continuum had not
changed significantly over this time, so this decline represents a
real and dramatic decrease in the intensity of the emission line
\citep{jwfc01}.

Our new observations show that this long-term decline in brightness of
the emission line has not continued; the EW has recovered to levels
last seen in 1999--2000, when the X-ray emission seen by \textit{RXTE}
was at its peak.


\subsubsection{H$\alpha$ morphology}

As we found in our previous observations \citep{jfw99,jwfc01}, the
H$\alpha$\ line shows multiple components, with a broad and a narrow
component at different velocities.  We fit Gaussian profiles to the
H$\alpha$ line using the {\tt specfit} package in {\sc iraf}
\citep{kri94}. After normalising each spectrum by a low-order
polynomial fit to the continuum, we fit either two or three gaussians
to the H$\alpha$\ line. Only spectrum C, taken on 2013 Feb 26 (phase
0.257) clearly required a third component to fit the line profile. The
other spectra are fitted with a broad component (with $\mathrm{FWHM}
\sim 1200\kms$) plus a single narrow component ($\mathrm{FWHM} \sim
500\kms$).

\begin{figure}
     \centerline{\psfig{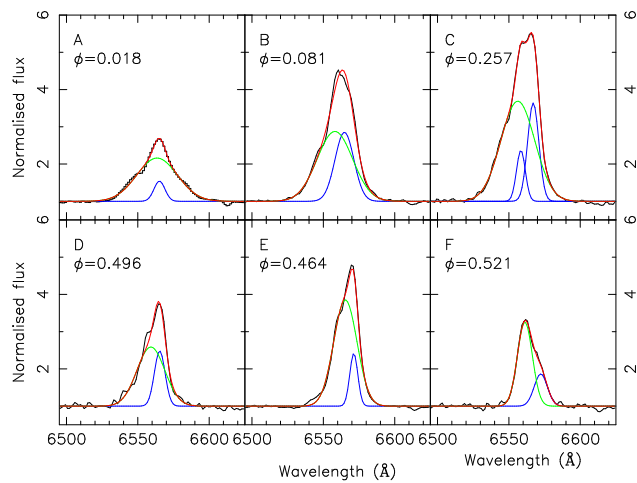}}
     \caption{Line profiles of H$\alpha$ in all six epochs; the name
     and phase of each observation is shown in the top left
     corner. The spectra were first normalised by a polynomial fit to
     the continuum. The line is broad and symmetric near phase 0, but
     at other phases the broad and narrow components have different
     velocities, with the broad component always blueward of the
     narrow component.  }\label{fig:Halpha}
\end{figure}

The fits to the H$\alpha$ profiles are shown in
Figure~\ref{fig:Halpha}.  Near phase 0 (panel A), the line is broad
and symmetric, with both broad and narrow component aligning at almost
the same velocity. At other phases, the two components appear at very
different velocities, with the narrow component (or components) always
redward of the broad component. This trend of line shape with phase is
similar to that seen in \citet{jwfc01}, where the line was observed to
change from single-peaked at phase 0, to double-peaked for phases
0.6--0.9. \citet{jwfc01} ascribed the appearance of double-peaked
lines during the second half of the orbit to the creation of a
Keplerian disk during the orbital cycle, a disk that is subsequently
destroyed during the next periastron passage.  Unfortunately, none of
our current spectra were taken during the second half of the orbit, so
we cannot check if this behaviour has persisted.


\subsubsection{He I lines}

The spectra also show emission lines of \hei\ $\lambda 6678$\ and
$\lambda 7065$. Spectra A--C show the line in emission; in spectra
D--F the emission line is much weaker, and there is also an absorption
component in both lines. 

(Further discussion of the emission lines is deferred to Johnston,
Wu \& Soria 2015\nocite{jws15}, in prep., where we will discuss
spectroscopic observations of the source over many years).



\subsection{Reddening}
\label{sec:reddening}

As discussed above, the extinction to the source is not well
constrained. We use our spectra to determine this extinction in
two different ways: using the diffuse interstellar bands, and using the
Balmer decrement.

\subsubsection{Diffuse interstellar bands}
\label{sec:DIBs}

\begin{figure}
     \centerline{\psfig{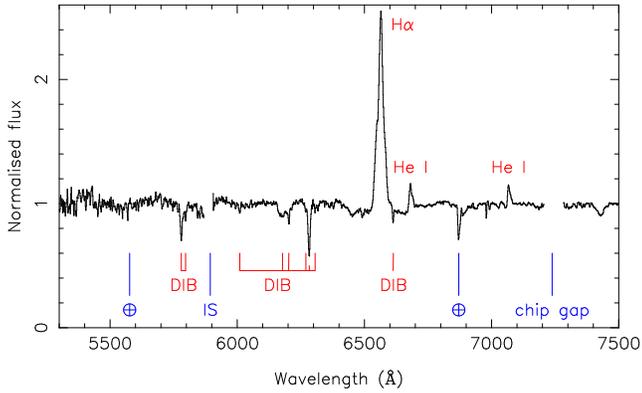}}
     \caption{Spectrum A in the region around H$\alpha$, with features
     identified. ``DIB'' indicates the diffuse interstellar bands at
     (from left to right) $\lambda \lambda$ 5780, 5797, 6010, 6178,
     6202, 6270, 6283, 6306 and 6613. Several terrestrial features
     (the [\ion{O}{i}] $\lambda$ 5577 line and the B band) are
     indicated; the gaps near 5900$\;$\AA\ and 7200$\;$\AA\ are due to
     interstellar Na D and the chip gap.  }\label{fig:DIBs}
\end{figure}

Figure~\ref{fig:DIBs} shows the normalised spectrum in the region
around H$\alpha$ with features identified. Apart from a couple of
terrestrial features, all but one of the observed absorption lines are
diffuse interstellar bands (DIBs). DIBs are absorption features of
unknown origin which are observed in the spectra of stars seen through
significant column densities of interstellar material \citep[see][for
a review]{her75}.

We can use the equivalent widths of DIBs to determine the extinction
$E(B-V)$, using the relations from \citet{her75}, who found a
correlation between line strength and colour excess for reddened
stars. We measured the strengths of several of the strongest DIBs in
our highest S/N spectrum, spectrum A: these measurements are shown in
Table~\ref{tab:DIBs}. \citet{her75} tabulated coefficients $a_0$ for
the correlation between equivalent width $W$ of 17 different DIBs and
extinction $E(B-V)$, of the form $W=a_0 E(B-V)$. We used these
tabulated $a_0$ values for three of the strongest DIBs
($\lambda\lambda$ 6010, 6202 and 6613) to estimate the colour excess
towards Cir X-1.

\begin{table}
\caption{Line strengths of several DIBs, and the reddening derived
from the relations of \citet{her75}. Column 2 shows the equivalent
width $W$\ in m\AA\ measured in Spectrum A; column 3 shows the
coefficient $a_0$\ from Herbig's Table 4, where $W=a_0 E(B-V)$; column
4 shows the derived colour excess $E(B-V)$. }.
\label{tab:DIBs}
\begin{tabular}{cccc}
\hline
     & $W$    &       &          \\ 
Line & (m\AA) & $a_0$ & $E(B-V)$ \\
\hline
6010 & 453 & 159 & 2.85 \\
6202 & 655 & 288 & 2.27 \\
6613 & 625 & 285 & 2.19 \\
\hline
\end{tabular}
\end{table}

The mean colour excess from these three lines is $E(B-V) = 2.44 \pm 0.21$,
which corresponds to an extinction of $A_V = 3.1 E(B-V) =
7.6 \pm 0.6\;\mathrm{mag}$.

\subsubsection{Balmer decrement}
\label{sec:balmer}

We can also estimate the reddening from the Balmer decrement, since
our spectra cover both H$\alpha$\ and H$\beta$. As seen in
Fig.~\ref{fig:spectrum}, the H$\alpha$ line is strong and broad, while
the H$\beta$ line is not detected. We use the prescription of
\citet{hob84} to estimate an upper limit to the flux from H$\beta$: we
set the $3\sigma$\ upper limit to be the product of three times the
rms fluctuation in the continuum and the line width $\Delta v$, which
we have assumed to be the same as the width of the H$\alpha$
line. Using spectrum A, which has the strongest H$\alpha$ line as well
as the highest S/N, we measure $f_\mathrm{H\alpha} = 5.3 \times
10^{-15}\;\mathrm{erg}\,\mathrm{cm}^{-2}\,\mathrm{s}^{-1}$, and
$f_\mathrm{H\beta} < 9.0 \times
10^{-17}\;\mathrm{erg}\,\mathrm{cm}^{-2}\,\mathrm{s}^{-1}$.  Hence we
have 
\[ 
\frac{f_\mathrm{H\alpha}}{f_\mathrm{H\beta}} > 65 
\] 
If we assume an intrinsic ratio for the Balmer-line intensities of
2.86 \citep[case B recombination; see][Table 4.1]{ost89}, then the
colour excess between H$\alpha$\ and H$\beta$\ is
\[ 
E(\mathrm{H}\beta - \mathrm{H}\alpha ) > -2.5 \log \left[
\frac{2.86}{(f_\mathrm{H\alpha}/f_\mathrm{H\beta})}\right] = 3.39 
\]
This can then be converted to the standard extinction $E(B-V)$ via
$E(B-V) = 0.863 \, E(\mathrm{H}\beta - \mathrm{H}\alpha)$ \citep[see
e.g. appendix of][]{mm72}, so $E(B-V) > 2.93$. We thus derive a lower
limit on the extinction $A_V = 3.1 \, E(B-V) > 9.1$, which is somewhat
larger than our estimate from the DIBs (Sect.~\ref{sec:DIBs}).

The intrinsic Balmer ratio could be significantly less than 2.86 if
the material emitting the line is optically thick. Many X-ray binaries
have values for H$\alpha$/H$\beta$ close to 2.7, indicating the lines
are optically thin \citep[e.g.][]{ssn+96,swj99,shc+04}. However, there
are exceptions, e.g. in GX339$-$4, where the Balmer decrement was
observed to be close to unity in the soft state \citep{rcl14}. Given
the absence of H$\beta$ in our spectra, a value for the intrinsic
ratio $f_\mathrm{H\alpha}/f_\mathrm{H\beta}=1.0$ would give an even
higher limit for the extinction, $A_V > 12.1$. Such a high value for
the extinction is not consistent with our estimate from the DIBs, or
from X-ray spectra, so we will take our best estimate for the
reddening to be in the range $A_V = 7.6$ to 9.1~mag.

\subsection{Continuum shape}
\label{sec:continuum-shape}

With no spectral features from the companion star, we are unable to
further constrain its spectral type using spectroscopy. The absence of
stellar features suggests the optical light is not being dominated by
the companion star; this is true in many low-mass X-ray binaries,
where the bulk of the light in the optical region comes from
re-processed X-ray radiation from an accretion disk and/or the heated
face of the companion. The imperfect flux calibration of our spectra
means we cannot fit the spectra with any degree of certainty.

\section{Discussion}
\label{sec:discussion}

\subsection{Irradiated companion model}
\label{sec:site}

Our light curve for Cir X-1 shows the visible light from the source
drops by 1.2~mag in 4 days (a quarter of an orbit), and is consistent
with that seen in 1989 by \citet{mon92}. This rapid change, together
with a spectrum devoid of stellar features, suggests that the optical
light is dominated by re-processed X-ray radiation. The modulation
indicates that the reprocessing material is not axisymmetric around
the X-ray source; in X-ray binaries this is typically an asymmetric
accretion disk, or the heated face of the companion star.

In \citet{jfw99}, we proposed a model for Cir X-1, where the binary
consists of a neutron star and an intermediate mass companion in a
highly eccentric orbit. During periastron passage, the companion star
overfills its Roche lobe, causing a transfer of mass at a
super-Eddington rate, which in turn drives a strong matter outflow.
After periastron, mass transfer from the companion ceases as the star
detaches from the critical Roche surface, but accretion continues at a
near-Eddington rate as the neutron star captures the residual matter
in its Roche lobe. An accretion disk gradually forms, so between
phases 0.5--0.9 there is steady accretion. The disk is then disrupted
by tidal forces during the following periastron passage. Support for
this model was provided in \citet{jwfc01}, where we twice detected
double-peaked emission lines, in spectra taken almost a year apart, at
phases 0.88 and 0.62.

If the accretion disk is destroyed during periastron passage, then the
disk cannot be the site of the optical emission. Here, we consider the
possibility that the variation we see in the visible light curve comes
entirely from the heated face of the companion star.  

A donor star of radius $R$ at a distance $d$ from the neutron star
will subtend a solid angle $\theta = \pi \left(\frac{R}{d}\right)^2$
as seen from the neutron star. Thus the amount of X-ray radiation
intercepted by the companion will depend on both its size and on the
shape of the orbit. The orbital period is well known, $P =
16.6\;\mathrm{d}$, and the primary is known to be a neutron star, so
we need to know the mass of the companion and the orbital
eccentricity.

As a starting point, consider first the model from \citet{jwfc01}:
take the mass of the companion $M_1 = 4\Msolar$ and
$e=0.8$, so the mass ratio $q = M_1/M_2 = 2.86$. The Roche lobe radius
of a star at the periastron of an eccentric orbit with semi-major axis
$a$ may be approximated as
\[
   R_{L,\mathrm{peri}} = a(1-e) \frac{ 0.49 q^{2/3} }{ 0.6 q^{2/3} + \ln(1+q^{1/3}) }
\]
derived from Eggleton's approximation for the size of Roche lobes
\citep[][see e.g. Sepinsky et al. 2007]{egg76}\nocite{swk07}.
Assuming the companion star fills its Roche lobe at periastron, then
at periastron it intercepts $\sim 6\%$ of the X-ray flux from the
neutron star. As it recedes in its orbit, this fraction drops, until
at apastron it intercepts only 0.07\% of the flux, or nearly two
orders of magnitude less. The rapid drop in the optical flux that we
observe after phase 0 represents the irradiated donor star getting
further away and hence being heated less.

We can use the decline in the optical flux to put limits on the shape
of the orbit. The more eccentric the orbit, the more rapidly the
companion star recedes after periastron, so the more rapidly the flux
should drop. We used the Eclipsing Light Curve code
\citep[{\sc elc};][]{oh00} to investigate how the light drops as a function
of orbital eccentricity. We do not have enough data to do a fit, but
we can test some indicative models. We started with the same model as
above: $P=16.6\;\mathrm{d}$, $M_1 = 4\Msolar$, $M_2 = 1.4\Msolar$. We
take the neutron star to be a point source of X-rays with luminosity
$L_X$. We have set the argument of periastron to be $90^\circ$ (i.e. the
neutron star is in front of the companion star at periastron; required
to get maximum heating at phase 0) and the orbital inclination $i$ to
be $80^\circ$. A high inclination angle is suggested by the presence
of X-ray dips, but our results are not very sensitive to the exact
value chosen. 

\begin{figure}
     \centerline{\psfig{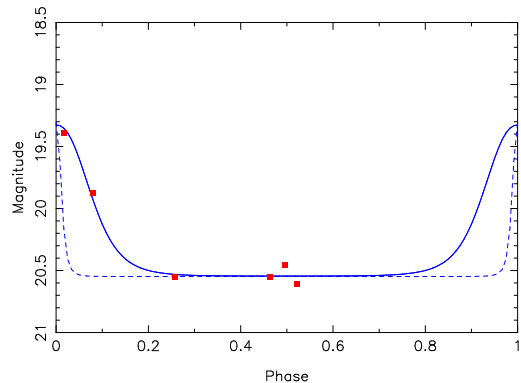}}
     \caption{Model from the {\sc elc} code of an irradiated star with $M_2
     = 4\Msolar$\ and $T=4500\;\mathrm{K}$. The red squares are our
     observed light curve (see Fig.~\ref{fig:Xray-lc}); the blue
     dashed line is a model with eccentricity $e=0.8$, while the solid
     line has $e=0.4$. Note that the X-ray luminosity is different in
     these two cases (see Table~\ref{tab:eccenfit}). The vertical axis
     is V-band magnitudes on an arbitrary scale.}\label{fig:ELC}
\end{figure}

We take the companion star to be a cool star ($T_1 =
4500\;\mathrm{K}$) with radius equal to the size of the Roche lobe at
periastron. For a given companion mass (and hence mass ratio $q$), the
amplitude of the variation is principally due to the X-ray luminosity
$L_X$: as the eccentricity increases, the two stars are closer
together at periastron, so a lower X-ray luminosity is required to
achieve the same change in temperature. However, the {\it width} of
the variation is due almost entirely to the eccentricity $e$: as the
eccentricity increases, the stars spend a smaller fraction of the
orbit at close distances with the companion facing us.
Table~\ref{tab:eccenfit} shows some typical values for our fitted
parameters.  
We find that an orbit with eccentricity $e=0.8$
\citep[the suggested value from][]{jfw99} does not match our light
curve well: the optical flux declines much too rapidly, taking only
$\Delta \phi = 0.04 = 16\;\mathrm{h}$ to drop by 1.2 magnitudes. The
eccentricity needs to be about 0.4 before it matches our light curve,
dropping 1.2 magnitudes in $\Delta \phi = 0.25 = 4\;\mathrm{d}$.
Figure~\ref{fig:ELC} shows the predicted light curve for these two
values of the orbital eccentricity, compared with our observed
values. A high value for the eccentricity is clearly ruled out.

\begin{table}
\caption{Example models fitted to the light curve. For assumed values
for the companion star mass and temperature, we look for the values of
the X-ray luminosity $L_X$ and eccentricity $e$ which reproduce the
shape of our light curve: a 1.2~mag drop in intensity in $\Delta \phi
= 0.25$ of an orbit.}\label{tab:eccenfit}
\begin{tabular}{cccc}
$M_1$       & $T_1$ & $\log L_X$ & $e$ \\
($\Msolar$) & (K)   & ($L_X$ in $\mathrm{erg}\,\mathrm{s}^{-1}$)    &     \\
\hline
4  & 4500 & 37.2 & 0.40 \\
4  & 7000 & 38.4 & 0.42 \\
10 & 4500 & 37.3 & 0.40 \\
10 & 7000 & 38.5 & 0.41 \\
\hline
\end{tabular}
\end{table}

Changing the mass of the donor star has very little effect on the
eccentricity required to fit the light curve. \citet{jnb07} modelled
the donor star as a $10\Msolar$ star, so we repeated the fit using
that mass. Again, we required an eccentricity $e=0.40$ to reproduce
the width of the drop in the light curve; the change in separation for
a different mass ratio is not nearly as important as the shape of the
orbit. Thus our light curve is insensitive to the mass of the donor
star.

The temperature of the companion, however, does have an effect. As the
temperature of the donor star increases, the X-ray luminosity required
to reproduce the observed 1.2-mag change in brightness also increases.
In going from $T=4500\;\mathrm{K}$ to $T=7000\;\mathrm{K}$, the X-ray
luminosity of the neutron star goes up by more than an order of
magnitude, from $\log L_X = 37.2$ to $\log L_X = 38.4$ (where $L_X$ is
the X-ray luminosity in $\mathrm{erg}\,\mathrm{s}^{-1}$).  An X-ray
luminosity $\log L_X = 37.2$ corresponds to $0.09 L_\mathrm{Edd}$ for
a $1.4\Msolar$\ star. The 2--20~keV MAXI light curve during the time
of our optical spectra (Fig.~\ref{fig:Xray-lc}) shows Cir X-1 had a
persistent X-ray flux within a factor of a few of this value (see
\S~\ref{sec:X-ray-lc}), so this level of X-ray illumination is not
unreasonable, particularly if we include the whole X-ray band. A
luminosity a factor of 10 higher, however, is not consistent with the
known source brightness during our observations. This favours the
companion being a cool star.

We assumed here that the donor star is filling its Roche lobe at
periastron. This is an upper limit to the size of the companion, and
sets a lower limit to the amount of X-ray heating required. If the
donor star is any smaller, it intercepts a smaller fraction of the
X-ray radiation and experiences less heating, and so requires a
greater X-ray luminosity to achieve the same brightness variation.

Thus our light curve can be explained by X-ray heating of the
companion, if the eccentricity of the orbit is not too high ($e \sim
0.4$) and the temperature of the companion star is low ($T \sim
4500\mathrm{K}$).  The mass of the companion star is not constrained
by the light curve. 

Many authors have assumed a rather high value for the eccentricity of
Cir X-1 ($e \sim 0.7$--0.9), based on modelling of radio
\citep{hlm80}, X-ray \citep{shi98} or optical data \citep{jfw99}. By
contrast, \citet{jnb07} modelled the system as a $10\Msolar$
supergiant B5--A0 star in an orbit with $e=0.45$. This eccentricity
and mass are compatible with our model, though the temperature of the
companion is not.

We also see strong and variable emission lines in our spectra, whose
shape and velocity change with orbital phase.  At phase 0, the
emission line is broad and symmetric, while at other phases the line
is asymmetric, with the narrow component always appearing redward of
the broad component.  
It is not clear where these components arise,
but they could well also be associated with the heated face of the
companion star.

\subsection{Constraints on the companion}
\label{sec:constraints}

Despite having not detected any stellar features in our spectra, our
observations nonetheless place quite stringent constraints on the
nature of the companion star.

\begin{enumerate}

\item {\it Absolute magnitude:} The apparent magnitude of the system
at minimum light is $\mathrm{V} = 21.46 \pm 0.02$ \citep{mon92}, which
represents an upper limit to the brightness of the companion -- any
contribution from a disk etc. would mean the companion was fainter
still. The best estimate for the distance to Circinus X-1 is
$9.4\;\mathrm{kpc}$ \citep{hbb+15}. Using our range of estimates for
$A_V = 7.6$--9.1, this apparent magnitude corresponds to an absolute
magnitude of $M_V=-0.7$ to $-2.2$, so the upper limit to the absolute
magnitude of the companion star is $M_V=-2.2$.

\item {\it Age:} The age of the companion must be the same age as the
star which went supernova. Assuming the minimum mass for a supernova
progenitor is $8\Msolar$, this corresponds to a lifetime $\tau \le 4
\times 10^7\;\mathrm{y}$ (see below). \citet{hsf+13} determined the
age of the binary to be $< 4600$~yr, so the companion star cannot be
older than the sum of these two times. Depending on its mass, this
places strict limits on its evolutionary state.

\item {\it Size:} Using our irradiated star model for the light curve,
the donor star is assumed to be filling its Roche lobe. Given the high
X-ray luminosity of the source, this seems a reasonable
assumption. 
If the star is not filling its Roche-lobe, then it must be
transferring material via a wind. However, the long-term average X-ray
luminosity is around $10^{37}\;\mathrm{erg}\,\mathrm{s}^{-1}$; if the
mass is inefficiently transferred in a wind, the donor star would need
to lose at least $10^{-6}\;\Msolar \,\mathrm{yr}^{-1}$ to reach this
luminosity \citep[see e.g.][\S 4.9]{fkr02}. Stars less massive than
$\sim 15\Msolar$ do not have winds this strong, and more massive stars
are ruled out by the luminosity criterion. 

From this size constraint, we can derive a limit on the average
density of the donor star. There is a well-known relationship between
the period and the density for Roche-lobe-filling stars:
\begin{equation}
   \langle \rho \rangle = \frac{107}{P_\mathrm{h}^2 (1-e)^3}\; \mathrm{g}\,\mathrm{cm}^{-3}
   \label{RL}
\end{equation}
where $P_\mathrm{h}$ is the orbital period in hours. This relation is
derived by combining Paczy{\'n}ski's approximation for the Roche-lobe
radius \citep{pac71} with Kepler's third law. However, it is in fact
only valid for $q < 0.8$; for larger values of $q$ (larger donor
masses), Paczy{\'n}ski's relation underestimates the volume of the
Roche lobe. This means that the real average density of the star will
be {\it less} than the density $\langle \rho \rangle$ estimated using
equation~(\ref{RL}). For the orbital period of Cir X-1, this gives us
a limit on the density of the companion: 
$\langle \rho \rangle <
0.003 \;\mathrm{g} \,\mathrm{cm}^{-3}$ for $e=0.4$, independent of the
mass of the star $M_1$.
\end{enumerate}

We can now try to find stars which satisfy the above criteria: $M_V >
-2.2$, $\langle \rho \rangle < 0.003 \;\mathrm{g} \,\mathrm{cm}^{-3}$,
and $\tau \le 4 \times 10^7\;\mathrm{y}$. We used the PARSEC stellar
isochrones \citep{bpl+12}, and plotted the average stellar density
versus absolute magnitude for different isochrones.

\begin{figure}
     \centerline{\psfig{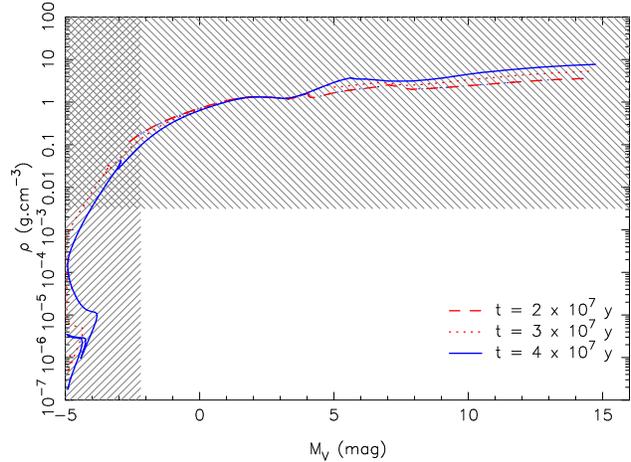}}
     \caption{Average density $\langle \rho \rangle$ vs. absolute
     magnitude $M_V$ for stars from the PARSEC isochrones. Three ages
     are shown: $\tau = 2 \times 10^7$ (red dashed line), $3 \times
     10^7$ (red dotted line) and $4 \times 10^7$~y (blue solid
     line). The last age is the upper limit to the lifetime of the
     star which produced the supernova, and hence of the age of the
     companion. The shaded regions show the regions disallowed by the
     constraints from our data: $M_V > -2.2$ and $\langle \rho \rangle
     < 0.003 \;\mathrm{g} \,\mathrm{cm}^{-3}$, so the companion star
     must fall in the blank area, where none of the models
     lie.}\label{fig:rhoV}
\end{figure}

Figure~\ref{fig:rhoV} shows three representative isochrones, ending
with $\tau = 4 \times 10^7$~y, which is the last model which contains
a $8\Msolar$ star, and hence represents the oldest possible age for
the companion star (constraint (ii)). The star must lie to the right
of $M_V = -2.2$ (constraint (i)), and below $\langle \rho \rangle =
0.003 \;\mathrm{g} \,\mathrm{cm}^{-3}$ (constraint (iii)). There are
{\it no} stars that satisfy all three constraints. Stars which have
low enough density to satisfy the density criterion are too bright to
satisfy the absolute magnitude criterion, unless they have lower mass,
in which case they do not satisfy the age criterion (as they have not
had time to evolve to large sizes and low densities).

\subsection{A companion star heated by the supernova}
\label{sec:heat}

What is the solution to this conundrum? We propose that it may lie in
the very young age of the binary since the supernova explosion. When
the supernova explodes, the impact of the ejecta on the companion star
can have a major, though comparatively short-term, effect. Several
authors have calculated the consequences for the companion star,
\citep[e.g.][]{mbf00,hy15}; they find that the companion is heated by
the supernova ejecta, suffers some mass loss, and will expand to have
higher luminosity and lower surface effective temperature. For
example, \citet{hy15} find models where the temperature decreases from
50,000~K to 4,000~K (their Fig. 8), depending on the scenario.  The
effect on the companion star will last a considerable time, of order
the thermal timescale of the star. \citet{prt12} and \citet{sks13}
find that the companion remains bright for several thousand years
after the explosion.

Thus we have the interesting situation that the donor star can be low
mass, and hence not have evolved, but also a giant star, due to the
fact that it is still recovering from the impact of the blast, less
than 5000~yr ago.  \citet{jnb07} concluded the companion was a hot
supergiant star in an orbit with $e=0.45$. We suggest that the star
does have a density similar to that of a supergiant, but much lower
temperature, due to it being ``puffed up'' by the deposition of energy
from the supernova. Their estimate of the eccentricity agrees with
what we find, $e \sim 0.4$.

There could also be another contribution to the heating of the
companion: tidal stresses along the eccentric orbit will provide
additional internal heat due to tidal friction, just as Io's interior
is kept heated by tidal forces from Jupiter.

Note that, if the star has been swelled up by the impact of the
supernova, it might not be necessary for it to fill its Roche
lobe. Red giant stars can have significant winds even at lower masses,
so the star might not completely fill its Roche lobe, with  wind
accretion, more focussed at periastron, providing the mass transfer.

We should note that we cannot rule out the possibility that the
secondary star is underdense because it has not yet reached the main
sequence. If the secondary has sufficiently low mass, it is possible
that the primary becomes a supernova before the secondary star has
contracted onto the main sequence. B stars with pre-main sequence
companions are known in wide orbits from the visual binary sample of
\citet{lin86}, so this possibility should be considered.

\subsection{Other possible scenarios}
\label{sec:other-models}

There are other possible scenarios for the system that might be able
to fit the observed constraints. If the optical radiation is not
coming from an irradiated companion, then the X-rays must be being
re-processed elsewhere in the system. An accretion disk is an obvious
possibility, though as described above we currently have no evidence
in our spectra for an ordered accretion disk existing at all phases of
the orbit. If the bulk of the light were coming from a disk, it would
explain the total absence of stellar features in our spectra. However,
in order to reproduce the strongly varying light curve, we would
require an asymmetric disk or accretion flow. 

We can also consider cases where it is not the companion star being
irradiated.  One possible model is that the companion is a Be star,
which is a rapidly rotating B star with an excretion disk (sometimes
called a ``decretion disk'', in reference to the fact that the mass is
flowing outwards instead of inwards; see \citealt{rcm13} for a
review). 
In this case, the X-rays are not produced by Roche-lobe
overflow from the star; instead, the baseline X-ray emission comes
from the wind from the disk, while the eccentric orbit means that near
periastron the disk reaches the Roche surface and produces outbursts. 
The varying optical light curve is explained by
irradiation from the central X-ray source, just as in our previous
scenario; both star and disk are more strongly irradiated at
periastron (with the disk probably contributing most of the irradiated
area). However, we do not prefer this model, for several reasons. Both
disk and central star are likely to be too hot to produce the 1.2-mag
drop (\S~\ref{sec:site}), plus the star is likely to contribute too
much light.  In addition, neither our optical spectra nor the infrared
spectra of \citet{jnb07} show double-peaked lines, which are typical
for Be stars, and the orbital period of Cir X-1 is shorter than any
other Be/X-ray binaries \citep[see e.g.][for a review]{rei11}.

\section{Conclusions}

We have found a strongly varying optical light curve for Cir X-1, with
the brightness changing by 1.2 magnitudes in four days (0.25 of an
orbit). This light curve is similar to that found by \citet{mon92} in
the 1980s, when the X-ray and radio brightness were much higher than
they currently are. We also observe strong, variable H$\alpha$ emission
lines, on top of an almost featureless but very red continuum.

We interpret the optical variability as arising from X-ray irradiation
of the companion star, or (less likely) a companion star plus disk, in
a moderately eccentric orbit ($e \sim 0.4$). If the star is filling
its Roche lobe, then the combination of age, brightness, and radius of
the companion star is incompatible with normal stellar models. The
very young age of the system -- determined by \citet{hbb+15} to be
less than 5000~yr since the supernova explosion -- gives us a possible
solution to this problem: the supernova explosion which produced the
neutron star deposited enough energy in the companion to ablate and
shock heat the star. The envelope of the star is puffed up, and the
star can take thousands of years to return to the main sequence. We
suggest that we are seeing the companion to Cir X-1 in this expanded
state.  A detailed model for the system will require simultaneous
X-ray and optical observations.

\section*{Acknowledgments}

We thank Dick Hunstead and Kinwah Wu for useful conversations, and the
anonymous referee for suggestions which greatly improved the paper.
Based on observations obtained at the Gemini Observatory (program ID
GS-2013A-Q-59, processed using the Gemini IRAF package), which is
operated by the Association of Universities for Research in Astronomy,
Inc., under a cooperative agreement with the NSF on behalf of the
Gemini partnership: the National Science Foundation (United States),
the National Research Council (Canada), CONICYT (Chile), the
Australian Research Council (Australia), Minist{\'e}rio da
Ci{\^e}ncia, Tecnologia e Inova\c{c}{\~a}o (Brazil) and Ministerio de
Ciencia, Tecnolog{\'\i}a e Innovaci{\'o}n Productiva (Argentina).

This research has made use of the MAXI data provided by RIKEN, JAXA
and the MAXI team. 

\bibliographystyle{mn2e}
\bibliography{strings,refs}

\end{document}